\documentclass{emulateapj}
\usepackage{graphicx}
\usepackage{epstopdf}

\usepackage[authordate1-4]{}
\tighten

\def\apjref#1;#2;#3;#4 {\par\pni\ #1,  #2, {\bf #3}, #4. \par}

\def\simlt{\lower.5ex\hbox{$\; \buildrel < \over \sim \;$}}
\def\simgt{\lower.5ex\hbox{$\; \buildrel > \over \sim \;$}}
\def\solar{\ifmmode _{\mathord\ odot}\else $_{\mathord\odot}$\fi}
\def\msun{\ifmmode {\rm M}_{\mathord\odot}\else $M_{\mathord\odot}$\fi}
\def\lsun{\ifmmode {\rm L}_{\mathord\odot}\else $L_{\mathord\odot}$\fi}
\def\xpvec{x\lla*p{\lower1ex\hbox{$\scriptstyle\sim$}}_{\scriptscriptstyle\perp}}
\def\xvec{x\llap{\lower1ex\hbox{$\scriptstyle\sim$}}}
\def\qvec{q\llap{\lower1ex\hbox{$\scriptstyle\sim$}}}
\def\to{\ifmmode \rightarrow\else $\rightarrow$\fi}
\def\Mc{M\raise0.5ex\hbox{c}}
\def\del{\nabla}

\def\none{\ifmmode ^{-1}\else $^{-1}$\fi}
\def\two{\ifmmode ^{2}\else $^{2}$\fi}
\def\ntwo{\ifmmode ^{-2}\else $^{-2}$\fi}
\def\three{\ifmmode ^{3}\else $^{3}$\fi}
\def\nthree{\ifmmode ^{-3}\else $^{-3}$\fi}
\def\four{\ifmmode ^{4}\else $^{4}$\fi}
\def\nfour{\ifmmode ^{-4}\else $^{-4}$\fi}
\def\five{\ifmmode ^{5}\else $^{5}$\fi}
\def\nfive{\ifmmode ^{-5}\else $^{-5}$\fi}

\def\g{\ifmmode {\rm g}\else g\fi}
\def\kg{\ifmmode {\rm kg}\else kg\fi}

\def\cm{\ifmmode {\rm cm}\else cm\fi}
\def\m{\ifmmode {\rm m}\else m\fi}
\def\km{\ifmmode {\rm km}\else km\fi}
\def\pc{\ifmmode {\rm pc}\else pc\fi}
\def\ly{\ifmmode {\rm ly}\else ly\fi}
\def\au{\ifmmode {\rm au}\else au\fi}

\def\s{\ifmmode {\rm s}\else s\fi}
\def\Hz{\ifmmode {\rm Hz}\else Hz\fi}
\def\y{\ifmmode {\rm y}\else y\fi}

\def\K{\ifmmode {\rm K}\else K\fi}
\def\ster{\ifmmode {\rm ster}\else ster\fi}
\def\erg{\ifmmode {\rm erg}\else erg\fi}
\def\dyn{\ifmmode {\rm dyn}\else dyn\fi}

\renewcommand{\baselinestretch}{1.0}\small\normalsize

\begin{document}

\title{Stellar Kinematics of Young Clusters in Turbulent Hydrodynamic Simulations}

\author{Stella S. R. Offner} \affil{Department of Physics, University of
California, Berkeley, CA 94720}  \affil{Center for Astrophysics, Cambridge, MA 02138 }
\email{soffner@cfa.harvard.edu}

\author{Charles E. Hansen}
\affil{ Department of Astronomy, University of California, Berkeley, CA 94720} 
\email{chansen@astro.berkeley.edu}

\author{Mark R. Krumholz}
\affil{ Department of Astronomy, University of California, Santa Cruz, CA 95064} 
\email{krumholz@ucolick.org}

\begin{abstract}
The kinematics of newly-formed star clusters are interesting both as a probe of the state of the gas clouds from which the stars form, and because they influence planet formation, stellar mass segregation, cluster disruption, and other processes controlled in part by dynamical interactions in young clusters. However, to date there have been no attempts to use simulations of star cluster formation to investigate how the kinematics of young stars change in response to variations in the properties of their parent molecular clouds. In this letter we report the results of turbulent self-gravitating simulations of cluster formation in which we consider both clouds in virial balance and those undergoing global collapse. We find that stars in these simulations generally have velocity dispersions smaller than that of the gas by a factor of $\sim 5$, independent of the dynamical state of the parent cloud, so that  subvirial stellar velocity dispersions arise naturally even in virialized molecular clouds. The simulated clusters also show
 large-scale stellar velocity gradients of $\sim0.2-2$ km s$^{-1}$ pc$^{-1}$ and strong correlations between the centroid velocities of stars and gas, both of which are observed in young clusters. We conclude that star clusters should display subvirial velocity dispersions, large-scale velocity gradients, and strong gas-star velocity correlations regardless of whether their parent clouds are in virial balance, and, conversely, that observations of these features cannot be used to infer the dynamical state of the parent gas clouds.
\end{abstract}

\keywords{ISM: clouds -- kinematics and dynamics-- stars:formation -- methods: numerical -- hydrodynamics -- turbulence }

\section{INTRODUCTION}

The majority of stars form in clusters of 100 or more members \citep{lada03}. The dynamics and distribution of stars within clusters have important ramifications for
planet formation, stellar multiplicity, mass segregation, cluster disruption, and any other property of stars and planets that can be influenced by interactions between stars in the dense environment of a young cluster.
For example, \citet{adams01} show that close encounters with neighboring stars can disrupt the outer parts of planetary systems (however, see \citealt{megeath08}). At an earlier stage, protoplanetary disks in dense clusters are subject to dynamical disruption (e.g.\ \citealt{spurzem09}), photoevaporation (e.g.\ \citealt{adams04}), and contamination with short-lived radio-nuclides \citep[e.g.][]{gounelle09, takigawa08, ouellette07}. On larger scales, radiation from neighboring stars in a cluster is an important feedback effect that can modify the stellar IMF \citep{adams08, offner09}, while gravitational interactions among stars lead to mass segregation \citep{bonnell98, mcmillan07, allison09}. All of these phenomena depend critically on the kinematics of newly formed stars, since this determines the frequency of star-star encounters. Kinematics also influence the fraction of star clusters that remain bound rather than dissolving into the field \citep[e.g.][]{baumgardt07}.

Observations of the kinematics of young stars are challenging because the stars spend much of their youth enshrouded by gas, making spectroscopy difficult. However, measurements have been possible in a few regions, most notably the Orion Nebula Cluster (ONC), a particularly interesting and well-studied young star cluster containing thousands of optically revealed stars as well as traces of remnant molecular gas. The ONC is observed to be centrally concentrated and mass segregated, with the most massive stars residing in the center (e.g.~\citealt{hillenbrand98}). Radial velocity measurements of stars within the ONC indicate a substantial stellar velocity gradient and significant correlation between the local gas and star velocities \citep{furesz08, tobin09}. As a result, \citet{tobin09} suggest that the ONC is kinematically young and not dynamically relaxed, which they interpret as requiring that the parent cloud be in a state of dynamical collapse \citep{hartmann01, bonnell03}, rather than evolving in quasi-equilibrum \citep{tan06a, kandt07, huff07}. \citet{proszkow09} use N-body simulations to demonstrate that collapsing clusters with subvirial initial conditions can indeed generate such observational signatures.

However, it is far from clear how to extrapolate from observations of stellar kinematics to the conditions within the host giant molecular cloud (GMC), particularly since not all components of a GMC share the same velocity dispersion. Instead, the dispersion of the line-of-sight (centroid) gas velocity decreases with density (e.g.~\citealt{blitz86, heyer01}). Gas velocity dispersions in prestellar cores are comparable to the sound speed and are much less than the global cloud virial velocity \citep{goodman93, andre07, kirk07, roso08}. Furthermore, the dispersion of the bulk velocities of the cores themselves is found to be subvirial \citep{andre07}, indicating the the relative motion between cores is 
small. Molecular gas in star forming regions such as Perseus and Ophiuchus also exhibits large-scale linear velocity gradients similar to those observed in the Orion stellar population \citep{andre07, kirk07, roso08}.  Perseus, in particular, resembles an elongated filament with a gradient of $\sim0.3$ km s$^{-1}$ pc$^{-1}$ along its length, similar to the ONC.

These observations of dense gas suggest that one cannot simply assume that, because stars are subvirial, the parent cloud must be too. Indeed, for gas the opposite appears to be the case. All of these subvirial signatures for dense components coexist with large-scale velocity dispersions that are consistent with virial balance \citep[e.g.][]{blitz07a, heyer08a, bolatto08a}. Simulations of turbulent, virialized clouds are able to reproduce the observed kinematics of dense gas \citep{padoan01, walsh04, Offner08b}, while simulations of clouds undergoing subvirial collapse have trouble producing quiescent cores like those we observe \citep{Offner08b}. Nor do clouds appear to convert large fractions of their mass into stars in a single dynamical time, as expected for a subvirial collapse  \citep[e.g.][]{klessen00, bonnell03}. Instead, observations both of individual star-forming regions \citep{evans09} and entire galaxies \citep{kandt07} show that, even for gas as dense as the cloud that formed the ONC, $\la 5\%$ of the mass per dynamical time can form stars.
 
The complicated relationship between the kinematics of dense and diffuse gas in molecular clouds suggests that it would be profitable to perform a systematic investigation of how the kinematics of newborn stars relate to the properties of the clouds from which they formed. Such an investigation is also of broader interest, since, as noted above, the kinematics of young clusters have a strong influence on planet formation, mass segregation, and other aspects of star and star cluster evolution. Thus it is important to develop first-principles predictions for the expected motion of newborn stars as a function of parent cloud properties, a subject that has not been investigated prior to this work. In this Letter, we use turbulent, self-gravitating, adaptive mesh refinement (AMR) hydrodynamics simulations of star cluster formation to study the properties of the star clusters that form in clouds in a variety of dynamical states.  In particular, we examine the stellar velocity dispersion, projected velocities, and velocity gradients. We compare the simulation results with recent observations of the ONC and discuss the relevance to current modeling of star clusters. In \S \ref{parameters} we describe our simulation methods and parameters. In  \S \ref{results} we present our results, and we summarize our findings in \S \ref{conclusions}.

\section{SIMULATION PARAMETERS} \label{parameters}

For our study, we perform three simulations of an isothermal, non-magnetized molecular cloud using the ORION AMR code (e.g.~\citealt{truelove98, klein99}). We reexamine the driven and decaying turbulent simulations described in \citet{Offner08a}, and we perform an additional simulation with higher resolution and different turbulent conditions. 

We generate the initial state of the runs by driving turbulent motions in the gas for two crossing times without gravity. At $t=0$, a turbulent steady state has been achieved and the gravitational energy is comparable to the turbulent energy in the box. The cloud virial parameter is then 
\begin{equation}
\alpha = \frac{5 \sigma^2R}{GM } = 1.67
\end{equation}
\citep{bertoldi92}.
In the simulation we refer to as U1 energy injection is halted after the initial driving phase and the turbulence decays
on a dynamical timescale, initiating a global collapse of the type posited by \citet{tobin09}.
In the simulation D1 we continue turbulent driving so that the Mach number and virial parameter are approximately constant;
this is a very rough representation of the behavior expected in a cloud where feedback drives turbulence and prevents global collapse \citep[e.g.][]{li06, nakamura07}. 
The third simulation, D2, begins with the same physical initial conditions, but we adopt a driving pattern with $k=1..2$ and drive the turbulence at a constant energy injection rate. As gravitational motions become significant, the net kinetic energy in the box increases and the virial parameter rises. 
The parameters for the three runs are summarized in Table \ref{table1}. 

Isothermal self-gravitating gas is scale free, so it is possible to scale the simulation values to observed cloud properties using the characteristic Jeans length and Mach number (see scaling relations in \citealt{Offner08b}). However, the main quantity of interest, the ratio of the stellar dispersion to the gas dispersion, is independent of the physical values we adopt. To estimate physical stellar velocities, we adopt a gas temperature of $T=20$ K and  a gas density of $\rho = 9.74\times 10^{-21}$ g cm$^3$, which are similar to the conditions observed in Orion \citep{johnstone01, johnstone06}. This corresponds to a box length of 2 pc and a cloud mass of 1150 $\msun$. 

We run the simulations with gravity for
one mean-density
freefall time, $t_{\rm ff}$.  We introduce a sink particle whenever the Jeans conditions is exceeded on the finest AMR level \citep{krumholz04}. These particles represent stars, serving as an indicator of the stellar cluster velocity dispersion  imparted by the initial turbulent conditions. 

Our simulations neglect radiative feedback, which likely has negligible influence on the initial stellar velocity. The simulations also neglect magnetic fields, which may affect the details of the simulated turbulence and exact distribution of gas densities. However, strong magnetic fields are more likely to further constrain the motion of the dense gas and forming stars rather than enhance it (relative to non-magnetized gas).

\begin{deluxetable}{ccccccc}
\tablewidth{0pt}
\tablecolumns{7}
\tablecaption{Simulation Parameters\tablenotemark{a} \label{table1}}
\tablehead{  
\colhead{} & \colhead{$N_0^3$\tablenotemark{b}} & \colhead{$\Delta x_{\rm min}$\tablenotemark{c}} & \colhead{Driving Method}  & \colhead{$k_{\rm min}..k_{\rm max}$}& ${\mathcal M}_f$\tablenotemark{d} & $\alpha_f$\tablenotemark{e}}
\startdata
U1 & 128$^3$	& 200 & No Driving  & 3..4 & 3.9 &  1.1 \\
D1 & 128$^3$	& 200  & Constant $\mathcal{M}$ & 3..4  & 8.5  & 1.7 \\
D2 & 512$^3$	& 100  &  Constant $dE/dt$ & 1..2 & 12.0& 3.4  \\
\enddata
\tablenotetext{a}{All simulations have the same initial $L_0$, $\rho_0$, $T_0$, ${\mathcal M}$. }
\tablenotetext{b} {Number of cells on the base grid.}
\tablenotetext{c} {Minimum cell size in AU.}
\tablenotetext{d} {Final 3D Mach number at $1t_{\rm ff}$.}
\tablenotetext{e} {Final virial parameter at $1t_{\rm ff}$.}
\end{deluxetable}

\section{RESULTS} \label{results}

In this section, we analyze the distribution and velocity information of the stars that form in each simulation. The simulation with decaying turbulence, U1, produces only 20 stars. D2 with the highest amount of turbulence and highest resolution forms 74 stars. In our simulations we do not resolve close binaries, and so in some cases the stellar properties represent the net mass and momentum of a stellar system. At the final time, the runs have efficiencies of $\sim$ 10-15 \%. 
Because we do not include protostellar winds, we neglect ejection of mass from prestellar cores, and thus likely overestimate both the stellar masses and the overall efficiency by a factor of 
$3-4$
\citep{enoch08, rathborne09}. However, this shift in stars' masses is unlikely to affect their kinematics.

\subsection{Stellar Velocity Dispersion}

\begin{figure}
\plotone{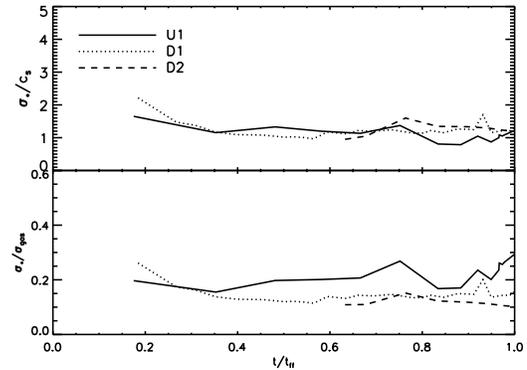}
\caption{The top plot shows the 3D star velocity dispersion as a function of time, while the bottom plot shows the dispersion normalized to the gas velocity dispersion.
\label{sigmavst}}
\end{figure}

Figure \ref{sigmavst} shows the 3D stellar velocity dispersion as a function of time.
The larger driving scale in D2 works to delay the onset of collapse so that stars form much later  than in the D1 and U1 calculations.
For U1 and D1, the stellar velocity dispersion is sonic to transonic. Once a sufficient number of stars are formed, the velocity dispersion trend appears to be mostly  uncorrelated with the simulation Mach number. In the undriven case, the stellar velocity dispersion declines slightly, while remaining nearly fixed in the simulation with constant Mach number driving. In D2, the dispersion decreases slightly with increasing Mach number.  
However, in all cases the stellar dispersion is significantly less than the global gas velocity dispersion:
$\sigma_*/\sigma_{\rm gas} \simeq 0.2$. This corresponds to final virial parameters of  
$\sim$ 0.10, 0.09 and 0.10 for U1, D1, and D2, respectively, using the total mass in stars and gas. If all the remaining gas were to disperse, leaving only stellar mass, then the virial parameter of the stars would be measured to be $\sim$ 0.8, 0.6, and 1.1 for the runs U1, D1, and D2, respectively.

This indicates that only if the initial gas cloud is significantly supervirial does the virial parameter of the stellar component become greater than unity, and that star clusters forming in turbulent virialized clouds naturally begin with subvirial velocities. Consequently, the subvirial nature of the stellar velocity dispersion is not necessarily an indication of collapse. 
Indeed, \citet{proszkow09} find that they are able to reproduce the kinematic signature of Orion assuming an initial virial ratio of $\sim 0.1$ (for stars plus gas), the value produced by simulation D1, in which the gas cloud is turbulent and virialized. Similarly, \citet{allison09}  assume an initial cluster virial parameter of 0.8, much less subvirial than we find for our simulated clouds with virial parameters of $\alpha \simeq 1$. 
If the ratio of the stellar dispersion to the gas dispersion influences how quickly the cluster undergoes mass segregation, then most clusters with substantially subvirial initial stellar velocities dispersions may segregate on short timescales. 

\subsection{Projected Stellar Velocities }

\begin{figure}
\plotone{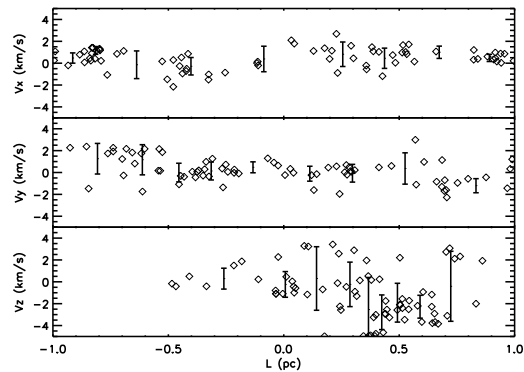}
\caption{The figure shows the projected line-of-sight velocities in the three cardinal directions as a function of linear position in the sky for the D2 run at $1t_{\rm ff}$. The error bars show the standard deviation of the velocities binned with a constant number of 10 per bin.
\label{plotvelbin}}
\end{figure}

In this section we investigate the stellar velocities projected along the line-of-sight to the cluster. For the most detailed examination, we use the final time of the D2 simulation, which has the best statistics. Figure \ref{plotvelbin} shows the velocities projected along the three cardinal sight-lines.  The velocities are somewhat clustered with only a couple instances of large deviations.  We have further binned the projected velocities into groups containing 9 stars as a function of position and we have overlaid error bars indicating the velocity dispersion in each bin. These plots look similar to Figure 9 in \citet{proszkow09} of the ONC, which shows evidence of a velocity gradient across the cluster.

\begin{figure}
\plotone{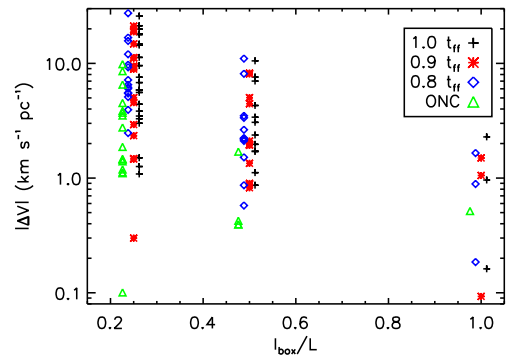}
\caption{The figure shows the distributions of gradients, $|\del V|$,  for sight lines along the x, y, and z directions as a function of box size. The ONC stellar data has been provided by \citet{tobin09}. The different datasets are offset slightly for clarity.
\label{grad}}
\end{figure}

We derive the stellar velocity gradient by fitting the line-of-sight velocities with the function $V = V_0 + \del V_x X+  \del V_y Y$. 
Although the data may not be exactly linear in nature (e.g. the ONC exhibits a linear gradient mainly in the cluster center), we can characterize the change in the velocities across the cloud.
Since we use periodic boundary conditions the velocity bins tend to wrap around the domain. Thus, we can better compare to the data by dividing each projection into smaller boxes and investigating the velocity trend in each box separately. 
In Figure \ref{grad}, we show the velocity gradients, $|\del V| = (\del V_x ^2 + \del V_y ^2)^{1/2}$, along each projection for the entire domain and for boxes half and a quarter of the domain. We include gradients at three different times and plot ONC data from \citet{tobin09} fitted in the same manner. To limit statistical fluctuations, we only include gradients fitted to 4 or more stars.
The full and half box fits have median velocity gradient magnitudes of 1 km s$^{-1}$ pc$^{-1}$ and 2.1 km s$^{-1}$ pc$^{-1}$, respectively. The magnitude of the gradients increases in both the simulations and the ONC for smaller boxes due to smaller statistics, where stars separated by several shock fronts may easily acquire significantly different velocities.
For comparison,  \citet{proszkow09} find that the ONC has a radial velocity gradient of 1.1 km s$^{-1}$ pc over half of its length, which corresponds to an overall gradient of $\sim$ 0.5 km s$^{-1}$ pc.
This value is similar to what we find for our simulated virialized clusters. Thus velocity gradients are not necessarily a hallmark of a subvirial turbulence or a strongly collapsing cloud.
Instead, large-scale velocity gradients appear to be a signature of the fact that both our simulations and real molecular clouds have most of their turbulent power in large-wavelength modes.

We find that the linear fits of the projected velocities fluctuate somewhat as a function of time, but they tend to fall within 0.2-2 km s$^{-1}$ pc$^{-1}$ on the largest scales. 
These values are similar to gradients in the gas centroid velocity observed in regions like Perseus and Ophiuchus \citep{andre07, roso08}.

\subsection{Gas Velocities }

Observations of clusters frequently find that the star and gas velocities are correlated (e.g., \citealt{furesz08}). We can reproduce such observations by making simulated observations of the gas in particular molecular tracers. Following the procedure of \citet{Offner08b}, we generate a position-position-velocity cube of the gas intensity in the $^{13}$CO (J = 1 $\rightarrow$ 0) molecular line along one line-of-sight in the D2 run. We solve for the line emission, assuming that the gas is in statistical equilibrium and that radiative pumping is negligible.\footnotemark
\footnotetext{We obtain the molecular data from the Leiden Atomic and Molecular database \citep{schoier05}.}
We adopt a cloud distance of 400 pc and 26'' beam size. To model the telescope resolution, we smear each velocity channel with the Gaussian beam. 

\begin{figure}
\plotone{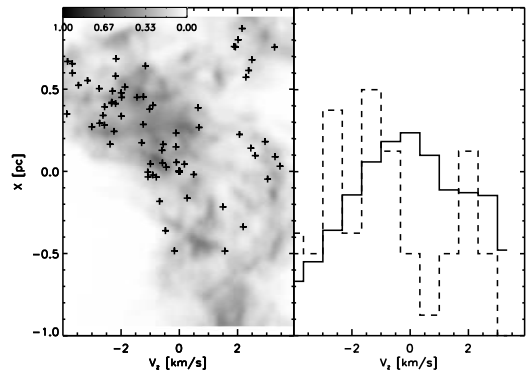}
\caption{The figure shows the position of the stars as a function of the line-of-sight velocity (crosses) in run D2. The gas velocity is overlaid, where the colorbar indicates the
relative  $^{13}$CO intensity averaged over the y coordinate.
 We have assumed that cloud is at a distance of 400 pc and is observed by a 26'' beam. The histograms on the right shows the distribution of gas (solid) and star (dashed) velocities normalized to the total intensity and star number, respectively. 
\label{gas_kin}}
\end{figure}

Figure \ref{gas_kin} shows the line-of-sight $^{13}$CO gas and star velocities;  
following Furesz et al.~(2008, their figure 10), to produce this  
figure from the 3-dimensional position-position-velocity cube, we have  
integrated along one direction on the plane of the sky, so the  
intensity shown in a given $(x, v)$ pixel is the sum over all $y$  
values for that $x$ and $v$. Our simulated image shows a strong  
correlation between gas and star velocities, similar to that seen in  
the star and gas velocity maps presented by \citet{furesz08}. For  
the projection shown in Figure \ref{gas_kin}, the mean stellar velocity is $\sim  
-0.8$ km s$^{-1}$, somewhat to the left of the mean $^{13}$CO gas  
velocity at 0.003 km s$^{-1}$. Most of the disagreement in the  
histogram occurs because not all regions traced by $^{13}$CO have  
formed stars. Projections along other directions show similar  
correlations between gas and star velocities.

The strong correlation between gas and star velocities might seem  
surprising, given that the gas and stars have such different velocity  
dispersions. The explanation is that the anti-correlation between  
density and velocity found in turbulent gas serves to artificially  
enhance the gas-star correlation. Figure \ref{gas_kin} is constructed by summing  
over the $y$ direction, so each pixel represents a density-weighted  
average velocity. The dispersion, however, is a density-weighted  
average of velocity squared, which, since velocity and density are  
anti-correlated, is much more heavily weighted toward low density,  
high velocity gas than what is shown in Figure \ref{gas_kin}. This low density gas  
is also that which is least correlated with the stars. The bias is  
exacerbated because $^{13}$CO is thermally excited only in relatively  
dense gas that is closely associated with forming stars. It is  
subthermally excited in the lower density gas that dominates the  
dispersion. Thus, both averaging and excitation serve to improve the  
correlation between $^{13}$CO emission and stellar velocities, while  
at the same time obscuring the low density gas that carries the bulk  
of the kinetic energy.

\section{CONCLUSIONS} \label{conclusions}

In this letter, we have investigated the kinematic properties of stars forming in simulated  turbulent self-gravitating clouds. These calculations represent the initial conditions of stars in clusters prior to dynamic evolution and cloud dispersal. We demonstrate that subvirial stellar velocity dispersions arise naturally from clouds in virial equilibrium. Thus, long lived clouds satisfying a quasi-steady state are not ruled out by observations of clusters with small stellar velocity dispersions. We find that gas virial parameters of order unity produce a star to gas velocity dispersion ratio of approximately 0.2.

We find that turbulent initial conditions easily generate stellar velocity gradients of the magnitude observed in star-forming regions.
Using a planar fit, 
we demonstrate typical gradients of 1 km s$^{-1}$ pc$^{-1}$, similar to the observed gradient in the ONC. 
We also show that for young clusters there is a strong similarity in the simulations between the dense gas traced by $^{13}$CO and the star velocities. 
Thus, trends in the stellar velocities as a function of position may be indicative of 
the dominance of large-scale turbulent modes,
rather than evidence of global collapse or cloud rotation as suggested by some authors. 

 \acknowledgments{
 Support for this work was provided by the US Department of Energy at the Lawrence Livermore National Laboratory under contract  B-542762 (S.~S.~R.~O., C.~E.~H.), 
the Alfred P.\ Sloan Foundation (M. R. K.), NASA/JPL through the Spitzer Theoretical Research Program (M. R. K.), the National Science Foundation through grants AST-0807739 (M.~R.~K.) and AST-0901055(S.~S.~R.~O). Computational resources were provided by the NSF San Diego Supercomputing Center through NPACI program grant UCB267.}


\end {document}